\pdfoutput=1
\documentclass[9pt,twocolumn,twoside]{osajnl}
\journal{preprint}
\setboolean{shortarticle}{False}

\newcommand{\figurewidth}{\columnwidth}

\title{Thermal-light heterodyne spectroscopy with frequency comb calibration}

\author[1,2,*]{Connor Fredrick}
\author[3]{Freja Olsen}
\author[3]{Ryan Terrien}
\author[4,5]{Suvrath Mahadevan}
\author[2]{Franklyn Quinlan}
\author[1,2,$\dagger$]{Scott A. Diddams}

\affil[1]{Department of Physics, University of Colorado Boulder, 440 UCB Boulder, CO 80309, USA}
\affil[2]{Time and Frequency Division, National Institute of Standards and Technology, 325 Broadway, Boulder, CO 80305, USA}
\affil[3]{Department of Physics and Astronomy, Carleton College, One North College Street, Northfield, MN 55057, USA}
\affil[4]{Department of Astronomy \& Astrophysics, 525 Davey Laboratory, The Pennsylvania State University, University Park, PA 16802, USA}
\affil[5]{Center for Exoplanets and Habitable Worlds, 525 Davey Laboratory, The Pennsylvania State University, University Park, PA 16802, USA}
\affil[*]{connor.fredrick@colorado.edu}
\affil[$\dagger$]{scott.diddams@nist.gov}

\begin{abstract}
Precision laser spectroscopy is key to many developments in atomic and molecular physics and the advancement of related technologies such as atomic clocks and sensors. However, in important spectroscopic scenarios, such as astronomy and remote sensing, the light is of thermal origin and interferometric or diffractive spectrometers typically replace laser spectroscopy. In this work, we employ laser-based heterodyne radiometry to measure incoherent light sources in the near-infrared and introduce techniques for absolute frequency calibration with a laser frequency comb. Measuring the solar continuum, we obtain a signal-to-noise ratio that matches the fundamental quantum-limited prediction given by the thermal photon distribution and our system's efficiency, bandwidth, and averaging time. With resolving power R${\sim}$10$^6$ we determine the center frequency of an iron line in the solar spectrum to sub-MHz absolute frequency uncertainty in under 10 minutes, a fractional precision 1/4000 the linewidth. Additionally, we propose concepts that take advantage of refractive beam shaping to decrease the effects of pointing instabilities by 100x, and of frequency comb multiplexing to increase data acquisition rates and spectral bandwidths by comparable factors. Taken together, our work brings the power of telecommunications photonics and the precision of frequency comb metrology to laser heterodyne radiometry, with implications for solar and astronomical spectroscopy, remote sensing, and precise Doppler velocimetry.
\end{abstract}

\begin{document}
\maketitle

\section{Introduction}

Optical frequencies are the most precisely measured physical quantities. The best laser spectroscopy has fractional uncertainty at the level of $1\times10^{-18}$~\cite{Beloy2021, Brewer2019}, and optical frequency combs~\cite{Diddams2020} have the capability to coherently synthesize and compare ratios of optical frequencies with uncertainty in the range of $10^{-21}$~\cite{Ma2007, Johnson2015, Yao2016}.  However, for many important optical and infrared systems of spectroscopic interest—such as those in astronomy and atmospheric remote sensing—the light to be analyzed does not come from a laser, but is thermal in origin. Here, one typically resorts to interferometric (Fourier transform) or diffractive (grating) spectrometers for spectral analysis. In such spectrometers, wavefront errors, instrument instabilities, and technical constraints on size limit the achievable precision as well as practicality. In this regard, heterodyne spectroscopy~\cite{Forrester1961} between a laser and a thermal source is a compelling option for making spectral measurements at known frequencies instead of at inferred wavelengths. Commonly called laser heterodyne radiometry (LHR), such thermal heterodyne has it roots in radio-astronomy~\cite{Sullivan1982, Parvitte2004} and has generally only been implemented for astronomy~\cite{Nieuwenhuijzen1970, Peyton1975, Kostiuk1983, Townes1984}, trace-gas sensing, and atmospheric spectroscopy~\cite{Ku1977, Menzies1977, Sonnabend2002} in the infrared region (e.g. 10 µm). However, with the ubiquity of high quality telecom fiber optic components there is growing interest in the use of LHR in the near-infrared region (1.5-1.7 µm) for remote detection and spectroscopy of gases in the earth’s atmosphere~\cite{Rodin2014, Wilson2014, Kurtz2016, Wang2020, Sappey2021}.

\begin{figure*}[t]
    \centering
    \includegraphics[width=\linewidth]{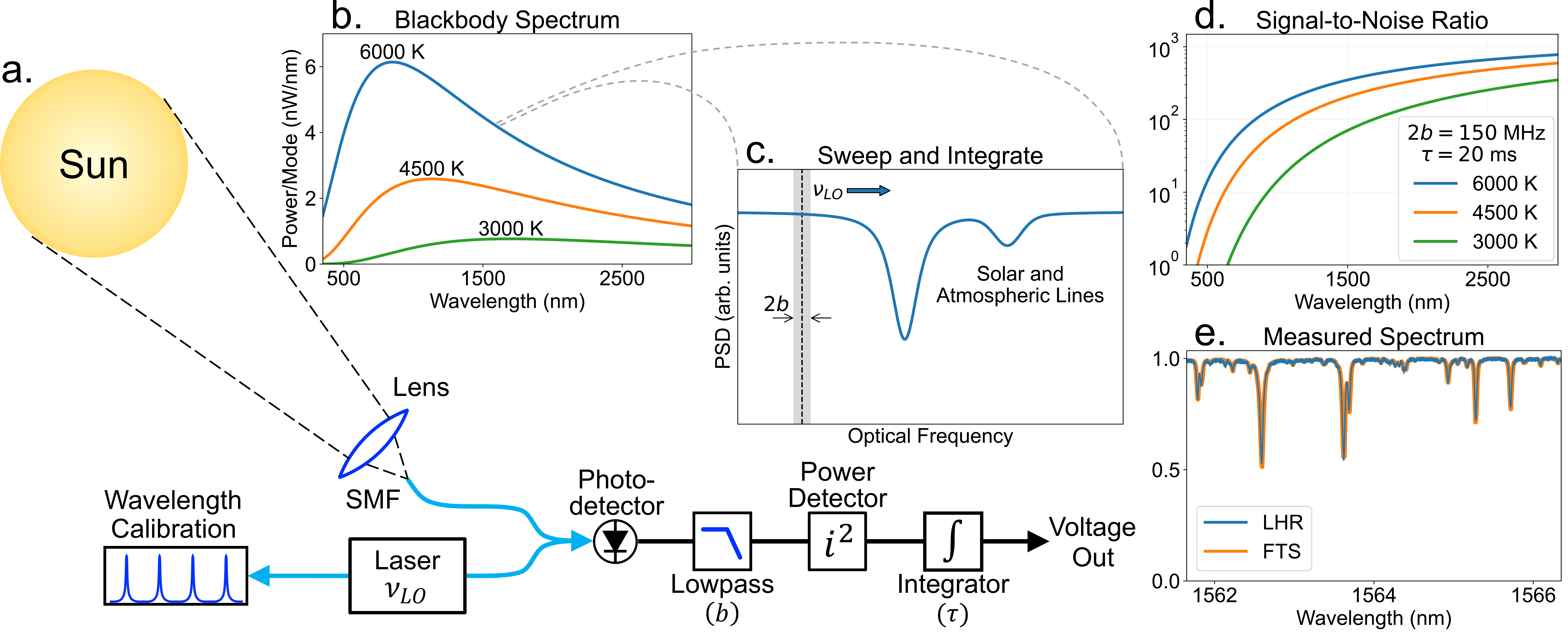}
    \caption{Laser heterodyne radiometry (LHR) concepts.
    \textbf{(a)}~Light from a thermal source, such as the Sun, is coupled into a single-mode optical fiber (SMF) where it is heterodyned with a laser and then processed with a few simple radio frequency (RF) components.
    \textbf{(b)}~The single-mode power spectral density (PSD) of a thermal source at various blackbody temperatures (solar surface is ${\sim}5800$~K).
    \textbf{(c)}~The local oscillator (LO) laser converts the spectral information from a small region about frequency $\nu_{LO}$ to the RF baseband. This information is sent through a low-pass filter (bandwidth $b$), power detector, and integrator (averaging time $\tau$) to provide a signal proportional to the power within the optical bandwidth $\Delta\nu = 2 b$. The full spectral profile can be reconstructed by scanning $\nu_{LO}$.
    \textbf{(d)}~The theoretical signal-to-noise ratio for effective bandwidth $2b=150$~MHz and averaging time $\tau=20$~ms.
    \textbf{(e)}~The solar spectrum near 1564~nm recorded with our LHR system and with a conventional Fourier-transform spectrometer~(FTS)~\cite{atlas1981, bass2000}.
}
    \label{fig:LHRConcept}
\end{figure*}

Figure~\ref{fig:LHRConcept} illustrates the operational concepts and parameters of a typical LHR system. Here, the source of interest is the Sun, but in general it can be any object that emits or reflects incoherent radiation. Through subsequent absorption and emission processes, either from atomic and molecular species in the Sun’s own photosphere or in earth’s atmosphere, spectroscopic information becomes imprinted on the thermal spectrum. A narrow linewidth local oscillator laser~(LO) is heterodyned with the thermal light and spectral information at difference frequencies with respect to the LO are mixed down into the radio frequency~(RF) domain. The RF power within an electrical bandwidth $b$ is proportional to the optical power in optical bandwidth $\Delta\nu = 2b$ centered on the LO frequency. This bandwidth sets the spectral resolution and can range from Hz to GHz using common radio-frequency electronics, which is a significant advantage of LHR. Without the complexity of large free-space delays or diffractive elements, LHR can achieve high spectral resolution in a compact apparatus and can be easily reconfigured by swapping out the RF filters. Scanning the LO frequency and recording the down-converted RF power reconstructs the optical spectrum.

In all such LHR measurements the stability of the LO frequency ultimately determines the spectral accuracy, and passive etalons, molecular absorption cells, or wavelength meters have been used to track the relative frequency of the LO~\cite{Rodin2014, Wilson2014, Kurtz2016, Wang2020, Sappey2021}. However, these frequency references are not tied to fundamental frequency standards, and will drift on their own or have unknown absolute accuracy. In this paper, we introduce laser frequency comb~(LFC) technology to LHR, thereby providing absolute frequency traceability and the capability of averaging or comparing spectral features over indefinite timescales. Our system achieves spectral resolving power of R=$\nu / \Delta\nu$=1,000,000 and is built on robust fiber-integrated lasers in the 1.5~µm region. Utilizing fiber-integrated laser power control and balanced photodetection, we largely remove technical noise and achieve sensitivity at the fundamental quantum noise limit~\cite{Zmuidzinas2003}.

We illustrate these advantages with LHR spectroscopy of laboratory, solar, and atmospheric sources. In studies of a solar iron line, we measure a signal-to-noise ratio (SNR) of about 300 with $20$~ms averaging time and $200$~MHz resolution bandwidth, a number which exactly agrees with the theoretical prediction. The full spectrum of the line is swept out in 10~s, and after averaging 60 spectra we are able to determine the line center to a precision less than 1~MHz, splitting the 3.7~GHz linewidth by a factor of $2\times10^{-4}$. Similar fractional precision is obtained on the HCN P28 line in a laboratory gas cell, with the narrower linewidth leading to an absolute frequency uncertainty of only 200~kHz. Crucially, these numbers are limited by the SNR of our experiment. The intrinsic frequency uncertainty of the instrument is still orders of magnitude lower and is ultimately only limited by the atomic frequency reference used to stabilize the frequency comb.

By removing the technical noise of the LO laser, the baseline drift, and the inaccuracy of the frequency axis we advance spectroscopy of thermal sources to a new level of precision and rigor, and have shown for the first time the capability of making spectroscopic LHR measurements that are truly limited by the fundamental quantum statistics of the observed thermal light. With quantum-limited sensitivity, absolute frequency calibration and improved long-term stability, our approach is transformative to a large body of work in atmospheric trace gas spectroscopy and solar astrophysics. High resolution measurements free of systematic noise are vital to the sophisticated modeling~\cite{Deng2021, Flores2021} employed for precise quantification of the distribution of greenhouse gases in Earth's atmospheric column~\cite{Weidmann2021}. In solar astrophysics, accurate knowledge of spectral line shapes and absolute frequency shifts can inform studies of helioseismology, magnetic activity, surface convection, and gravitational redshift~\cite{deming1986infrared, leone2003measuring, lohner2018convective, hernandez2020solar}. In addition, such precise measurements are essential for models of stellar magnetism and photospheric dynamics and the efforts aimed at disentangling such activity from center-of-mass Doppler shifts~\cite{Cegla2019stellaractivity}. The potential to fully resolve GHz-wide lines with sub-MHz absolute frequency precision across the near and mid-infrared (wherever stabilized comb light can be generated) opens new possibilities for high precision Solar and atmospheric LHR spectroscopy, which has previously been restricted to extremely narrow optical bandwidths surrounding isolated atomic or molecular transitions~\cite{goldstein1991absolute}. Modifications to this technique also hold the possibility to directly achieve cm/s level radial velocity precision (10-100 kHz) on the Sun, which has not been demonstrated to date. The ability to do so is a critical and open question relevant to the astronomical community searching for terrestrial-mass exoplanets in the habitable zones of nearby stars~\cite{crass2021extreme}. While LHR cannot replace traditional spectrographs for the radial velocity measurement of other stars~\cite{Mayor2003harps, Suvrath2012hpf, Schwab2016neid, Pepe2021espresso}, for detailed and high-resolution measurements of velocity and line shape on the Sun this technique is superior to other methods which at the level of a single line would struggle against instrumental systematics~\cite{Lemke2016solarfts, Cameron2019harpssun, Dumusque2018linebyline, Ninan2019h2rg}. Furthermore, we lay the groundwork for additional modalities of direct frequency comb spectroscopy of non-laser sources~\cite{Giorgetta2010, Boudreau2012}. This includes massively-parallel heterodyne with frequency combs, which could be employed not only for spectroscopy but also for long-baseline phased-array imaging in the mid-infrared~\cite{Hale2000, ireland2014dispersed}.

\subsection{Theoretical background}

The theory of LHR has been thoroughly covered by others~\cite{Abbas1976, Parvitte2004, Protopopov2014lhr}, and we simply state the important considerations for the system described in this paper. A single optical mode can only couple to a single statistical-mechanical thermal mode. The antenna theorem limits the product of the effective aperture~($A$) and field of view~($\Omega$) of an optical mode to the square of its wavelength~($\lambda$), $A \ \! \Omega=\lambda^2$~\cite{Siegman1966}. The power coupled into a single mode is maximized when the field of view is completely filled by the thermal source:
\begin{equation}
    \label{eq:power}
    P_\text{max} = h \nu \left< n \right> \Delta\nu
\end{equation}
where $h\nu$ is the photon energy, $\Delta\nu$ is the optical bandwidth, and $\left< n \right>$ is the mean photon occupancy given by the Planck or Bose-Einstein distribution, $\left< n \right>=\left(\exp\!\left[h\nu / kT \right] - 1\right)^{-1}$. Equation~\ref{eq:power} can also be viewed as the maximum power that can be delivered to a receiver's area over which the thermal source is unresolved and its light is spatially coherent. This expression, converted to a spectral density in terms of wavelength, is shown in Fig.~\ref{fig:LHRConcept}(b) for a few blackbody temperatures relevant to the Sun and other common stellar types. Note that at 1550 nm, there is only 5~pW within a resolution bandwidth of $\Delta \nu=150$ MHz for a thermal source with the temperature of the Sun (roughly 4~nW/nm single-mode power spectral density). The shape and extent of the optical mode is important for how information is retrieved spatially (see Section~\ref{sec:ideas}), but as long as the thermal source is larger than the field of view, improved imaging or tighter focusing onto the fiber will not change the coupled power. For standard single-mode fiber (SMF) at 1550~nm, only a simple collimating lens with mm scale aperture is required to achieve a field of view smaller than the ${\sim}0.5^\circ$ angular diameter of the Sun.

The heterodyne process is equivalent to passing the thermal light through a quantum-limited, phase-insensitive amplifier. If the LO shot noise dominates other sources of noise, this leads to the following signal-to-noise ratio~\cite{Zmuidzinas2003}:
\begin{equation}
    \text{SNR} = \frac{\eta {\left< n \right>}}{1 + \eta {\left< n \right>}} \sqrt{\Delta\nu \, \tau}
    \label{eq:SNR}
\end{equation}
where $\tau$ is the averaging time and $\eta$ is the probability that a photon emitted by the blackbody arrives at the detector and results in a photoelectron. The efficiency factor $\eta$ includes all transmission and coupling losses as well as the detector's quantum efficiency. The additional factor of $\eta {\left< n\right>}$ in the denominator is due to thermal photon bunching and has important implications. Regardless of the number of photons, or how hot the source is, the first term in Eq.~\ref{eq:SNR} is always less than 1. This means that useful signal-to-noise ratios can only be obtained with large optical bandwidths or averaging times. The bunching term is also why LHR has typically only been considered for the mid- and far-infrared, as starting with a relatively large number of photons gives an SNR that is both larger and less sensitive to optical loss. The signal-to-noise ratio rapidly drops off at short wavelengths due to reduced photon counts, but as seen in Fig.~\ref{fig:LHRConcept}(d), the SNR can assume significant values in the near-infrared for the reasonable experimental parameters that we employ.

\section{Experimental Setup}
\label{sec:setup}

\begin{figure*}[t!]
    \centering
    \includegraphics[width=\linewidth]{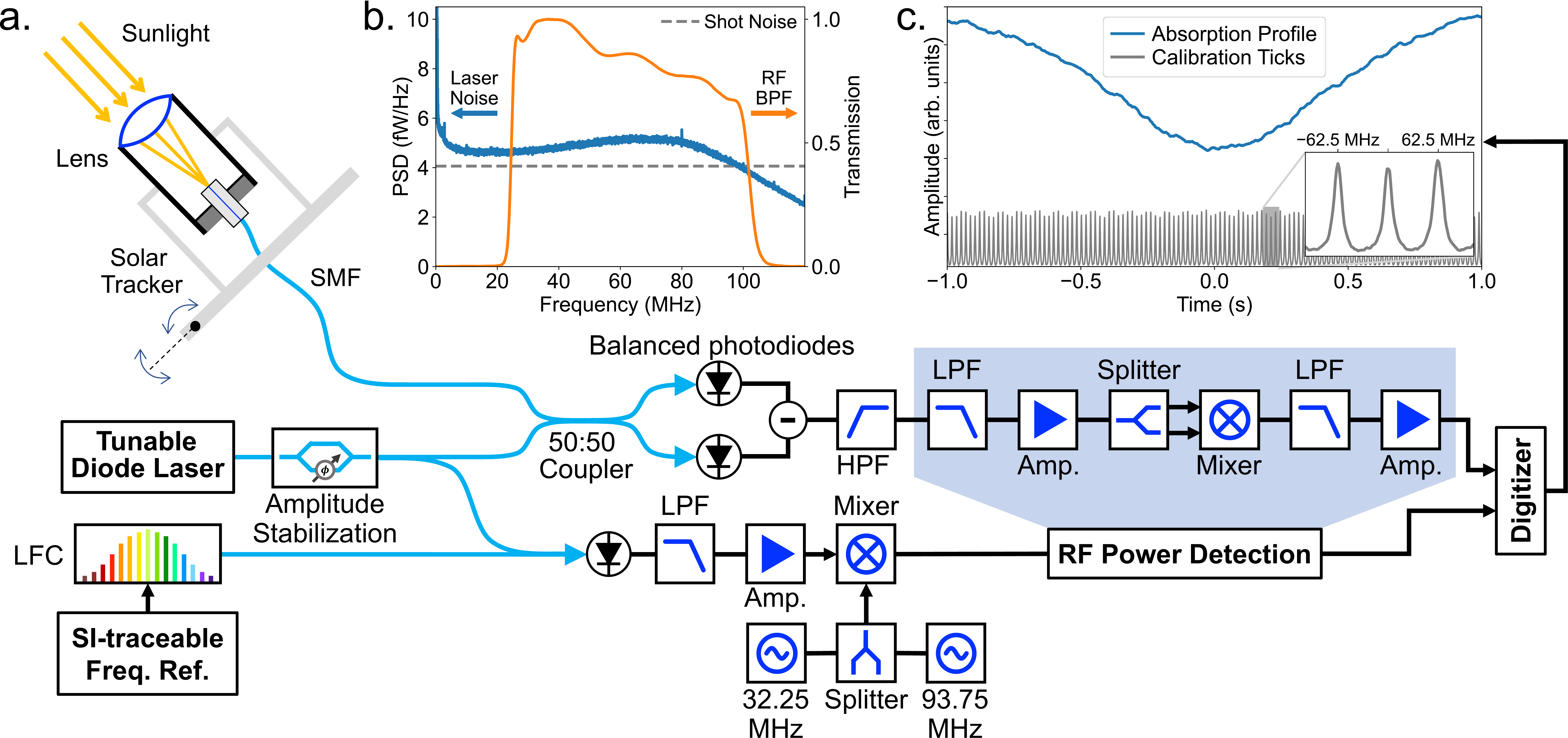}
    \caption{
    Experimental setup for laser-frequency-comb (LFC) calibrated LHR spectroscopy.
    \textbf{(a)}~Schematic of the key components and signal pathways, see text for details.
    \textbf{(b)}~Laser noise and LHR passband. The noise spectrum \textit{(blue)} from the balanced detector is within 1~dB of the calculated shot-noise limit \textit{(grey)}. The roll off at high frequency is due to the 100~MHz detector bandwidth. The effective RF bandpass filter \textit{(orange)} defines the doubled-sided optical passband centered on the LO frequency. The low-pass filter at 100~MHz sets a resolving power $R{=}\nu / \Delta\nu$ of approximately $10^6$ in the 1550~nm region. The high-pass filter at 25~MHz attenuates excess low-frequency technical noise.
    \textbf{(c)}~Simultaneous solar and calibration signals as recorded on an oscilloscope. Each calibration tick is separated by 62.5~MHz and is traceable to a stabilized LFC, allowing for absolute frequency calibration.
}
    \label{fig:setup}
\end{figure*}

Our experimental setup is depicted in Fig.~\ref{fig:setup}. Sunlight is coupled into a single mode fiber by a fiber collimator (Thorlabs 50-1550) mounted to a solar tracker (Orion Solar StarSeeker). The divergence of the collimator's Gaussian mode gives a $1/e^2$ angular radius roughly half that of the Sun's, so about $25\%$ of the solar disk is within the effective field of view. The SMF runs ${\sim}100$~m to an indoor laboratory where the LHR components are located. The sunlight is combined with a single-frequency diode laser in a 50:50 fiber coupler and then interfered on balanced photodetectors. Balanced detection allows measuring all of the collected sunlight while simultaneously rejecting amplitude noise associated with the laser~\cite{Collett1987Quantum}. Before the coupler, the laser is first passed through an amplitude stabilizer implemented with a fiber-coupled Mach-Zehnder intensity modulator. This servos the laser to a constant power, stabilizing the background level of the LHR signal as the laser frequency is swept. The servo maintains the sum of the two balanced detector outputs at 9.75~V, or at about 0.975~mW of total optical power.

The heterodyne signal is filtered and amplified before the RF power is detected. A 25~MHz high-pass filter removes low frequency technical noise due to the laser and detector electronics. A 100~MHz low-pass filter sets a sharp high frequency cutoff and gives an optical bandwidth of 200~MHz. The laser noise spectrum and the shape of the effective bandpass filter are plotted in Fig.~\ref{fig:setup}(b). Across the entire passband, the noise spectrum is within 1~dB of the calculated shot-noise limit assuming a quantum efficiency of 0.82 and a transimpedance gain of 25~V/mA. The shape of the effective bandpass filter was measured by recording the response of the power detection electronics to a swept RF tone generated by a synthesizer. Weighting and integrating the filter profile by the noise spectrum gives a double-sided equivalent noise bandwidth of 150~MHz. While the high pass filter at 25~MHz does create a non-ideal instrument response function with a 50~MHz gap in the center, excess noise below 25~MHz results in a smaller SNR when the filter is removed. In principle, with lower noise laser and photodetection electronics, the central gap in the detection band could be reduced to only a few hundred kHz using readily available RF components.

The RF power detection circuit utilizes the diode bridge of a double balanced mixer for rectification. A $0^{\circ}$ power splitter sends the amplified heterodyne signal to both input ports of the mixer. The output of the mixer is terminated into 50~Ohms and the DC voltage is used as a measure of the RF power. A final low-pass filter and preamplifier are applied before that signal is sent to the digitizer. The low-pass filter is a programmable second order RC filter and is set to give an effective averaging time of 21~ms and 210~ms for the solar and HCN measurements. The transfer function from broadband RF noise power at the input to DC voltage at the output was determined by heterodyning the LO with an amplified spontaneous emission (ASE) source and is nearly linear for the optical power levels involved. After frequency axis calibration, the transfer function is used to offset and scale the raw voltage signal so that it corresponds to RF noise power.

\subsection{Absolute Frequency Calibration}
Using the same LHR techniques as for the thermal light, the spectrum of a laser frequency comb~\cite{Diddams2020} is also mapped out through heterodyne with the LO laser. Our LFC is referenced to a NIST-calibrated hydrogen maser and provides a grid traceable to the SI second, with uncertainty of a few parts in $10^{13}$ or better on timescales from seconds to years ($\sim$20 Hz frequency precision at 1550~nm). This precise and accurate reference has been absent from other LHR measurements, and we believe it will be key to the long-term averaging and day-to-day repeatability required to meaningfully track the center frequency and line shape of solar or atmospheric lines.

The position of each comb tooth is given by the comb equation, which is defined by two degrees of freedom and an integer comb line number, $\nu_n = f_0 + n \, f_R$. When the offset frequency~($f_0$) and the repetition rate~($f_R$) are properly stabilized, knowledge of the comb line number~($n$) allows for absolute frequency calibration. For single, fixed frequency applications, $n$ is typically determined by a lower-resolution absolute measurement (e.g. with a wavelength meter) or by performing multiple heterodynes with the comb set at different repetition rates. However, since the frequency of our LO laser is continuously scanning those methods are not conveniently applied. We use the centroid of an absorption line from a reference gas cell to determine the integer $n$, but such determination could also easily be accomplished by separately recording the heterodyne between the scanning laser and a secondary, fixed laser locked to a known comb line. 

Since our LFC only has a comb tooth every 250~MHz, the heterodyne beat is first mixed with additional synthesizers at 31.25~MHz and 93.75~MHz to give a denser grid of calibration ticks~\cite{Jennings2017}. Using hardware identical to that in the thermal LHR arm (shown in the blue shaded region of figure~\ref{fig:setup}) but with 4~MHz double-sided bandwidth and 0.6~ms averaging time, RF power detection provides a voltage output every $f_{R}/4=62.5$~MHz for each calibration line. The thermal and the calibration spectra are recorded simultaneously on an oscilloscope, a sample of which is shown in~\ref{fig:setup}(c).

The calibration procedure consists of fitting Gaussian profiles to the calibration ticks to find their center and then linearly interpolating between those points to construct a time-to-frequency calibration curve. Since these are continuous time-domain measurements, the absolute frequency calibration is susceptible to the product of the sweep rate and the relative time delays between the two arms. The largest contributor to such delays are the final, time-averaging low-pass filters. Given a nominal sweep rate of 5~GHz/s, a 10~ms delay due to a 20~ms averaging time results in a 50~MHz frequency offset. The actual delays for each filter configuration were determined by splitting a slow sine wave, sending it into both arms' filters, and then measuring their relative delay at the oscilloscope. The sweep rate is automatically calculated as part of the time-to-frequency calibration curve, so once the fixed delays are known the effective frequency shift can be removed in post processing. However, this issue could be avoided altogether by sweeping in a step and hold pattern, such that the frequency of the LO is constant during the measurement of each spectral element. In that case, instead of fitting to and interpolating between the comb lines in the measured power spectrum, one could determine the LO frequency by directly counting the frequency of the heterodyne beat note or by servoing it to a prescribed value.

To estimate an upper limit on the sweep-to-sweep calibration uncertainty we injected LFC light into both arms of the LHR system. The frequency calibration procedure was applied using data from the first arm while cross-correlation was used on data from the second to determine the trace-to-trace frequency shift. For 5~s long traces spanning a 15~GHz optical window, this comb versus comb measurement yielded a frequency uncertainty at the few 10’s of kHz level, approaching the 1~kHz level after 1000~s of averaging.

Due to the additional RF processing, the absolute frequency of a spectral line measured by our LHR system is given by a modified comb equation:
\begin{equation}
\label{eq:comb}
\nu = f_0 + \left(n + \frac{1}{2}\right)\frac{f_R}{4} + f_c - f_{dly}
\end{equation}
where $f_0$ and $f_R$ are the maser-referenced carrier-envelope offset frequency and repetition rate of the frequency comb, $f_{dly}$ is the effective frequency axis shift due to the time delay between the two arms, and $f_c$ is the line center frequency relative to the calibration tick given by integer index $n$. The line center frequency is found by fitting a Voigt profile to the spectral line being measured. Electronic mixing has reduced the effective repetition rate and increased the number of calibration ticks by a factor of 4. The delay frequency shift ($f_{dly}$) is calculated from the scan rate of the LO laser and the time delay of the averaging low-pass filters, which were measured to effective frequency precisions 2 and 4 times better than the minimum given by the Allan deviations of the solar and HCN measurements. With stability near $10^{-13}$ for the offset frequency and repetition rate, the largest source of uncertainty is the determination of the line center frequency ($f_c$), which depends heavily on the SNR and line shape.

\section{System Characterization}

\begin{figure}[tb!]
    \centering
    \includegraphics[width=\figurewidth]{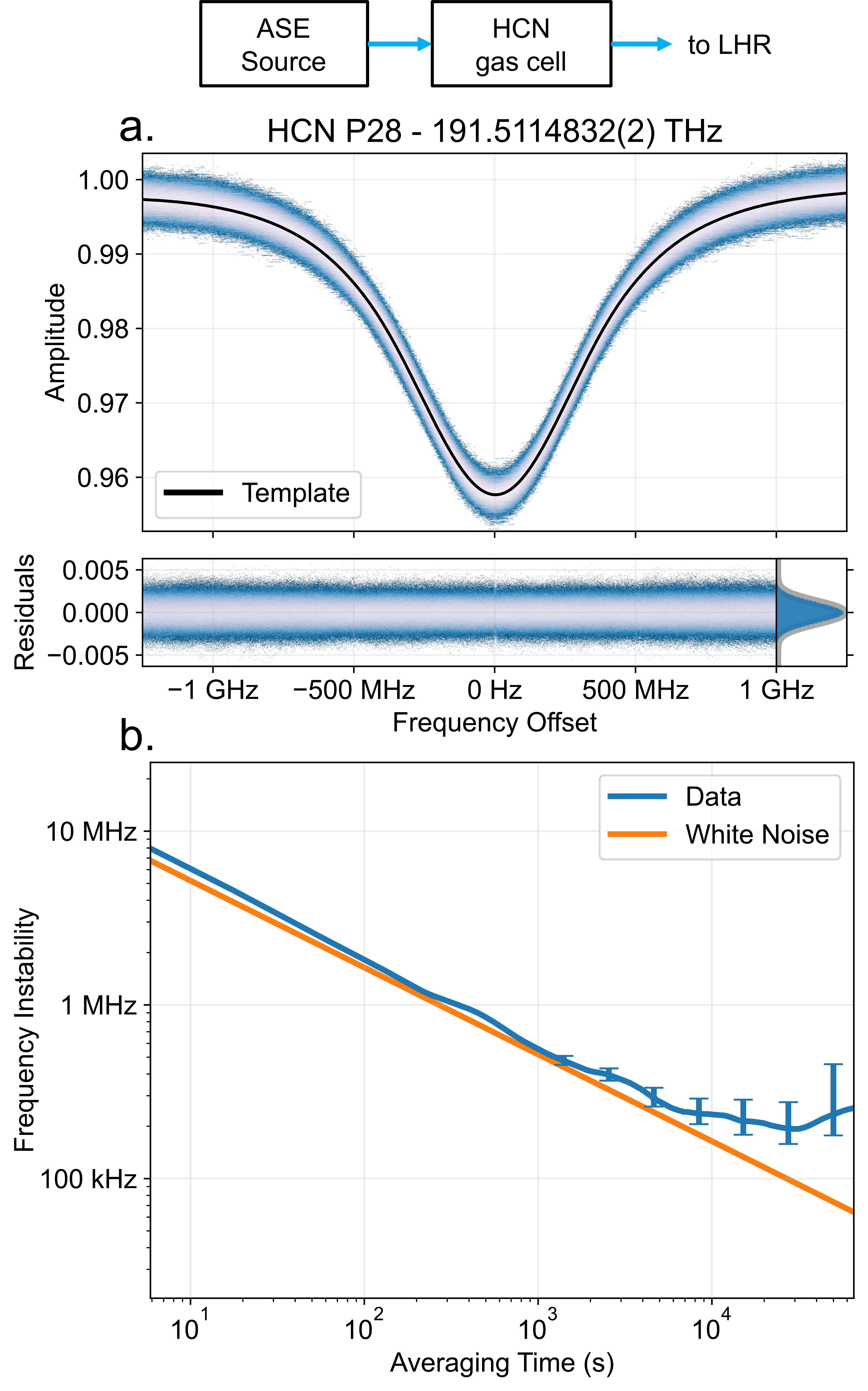}
    \caption{An H$^{13}$C$^{14}$N reference cell~\cite{srm2519a} is illuminated by amplified spontaneous emission (ASE). 
    \textbf{(a)}~LHR of the P28 line of HCN near 1565.40~nm~\cite{Swann2005hcn}. The data represents all 22,281, 6~s traces taken over the course of 36~hours. The amplitude of the transmission spectrum through the gas cell has been normalized to the continuum level of the ASE and the color is scaled to the density of points. The residuals with respect to the spectral template are plotted below. The inset at the bottom right shows the histogram of the residuals outlined by a Gaussian distribution of the same width in grey.
    \textbf{(b)}~The Allan deviation of the retrieved line centers and the theoretical white-amplitude-noise limit. The error bars represent the $\pm 1 \sigma$~confidence intervals.}
    \label{fig:HCN}
\end{figure}

We conducted LHR experiments using an amplified spontaneous emission source in the lab to explore systematics independent of instabilities due to telescope tracking, relative Solar velocity and rotation, as well as variable Solar flux (e.g. due to cloud cover). The broadband light from an unseeded semiconductor optical amplifier was attenuated to approximate the power spectral density of sunlight. Absorption lines imparted on the ASE by an HCN gas cell (NIST SRM~2519a~\cite{srm2519a}) served as stable spectral features to measure with the LHR setup. The results from a series of scans of the P28 transition spanning 36 hours are shown in Fig.~\ref{fig:HCN}. Due to the weak nature of this line and the short optical path length within the cell, the absorption profile only dips about 4\% away from the relative continuum level. The scan range of this measurement was restricted to a 4~GHz region around the P28 transition. Each scan was collected in 6~s with a resolution bandwidth of 200~MHz and an effective averaging time of 210~ms.

The amplitude and frequency noise are analyzed by comparing the individual traces against a template formed from a filtered average of the entire dataset (see Supplement for details). The two-dimensional histogram of the data from the individual traces is plotted against the spectral template in Fig.~\ref{fig:HCN}(a). The curves are normalized to the same mean amplitude, so the residuals solely represent the relative amplitude noise. The distribution of the residuals is Gaussian, as expected of amplitude fluctuations due primarily to shot noise.

The effective line center shift of each trace is calculated by finding the frequency shift that maximizes the trace's cross-correlation with the template. The Allan deviation of the frequency shifts yields an 8~MHz uncertainty for the 6~s traces, which reduces down to about 200~kHz after $10^4$~s of averaging. We compare the measured frequency uncertainty to an estimate of the extracted center frequency's sensitivity to white amplitude noise, calculated from the line shape and the statistics of the amplitude residuals (described in Supplement). As seen in Fig.~\ref{fig:HCN}(b), over the first few decades the frequency instability of the HCN line agrees with this white-amplitude-noise limit. This indicates that random amplitude fluctuations dominate the short and medium term frequency instability and the measurement is limited purely by the SNR.

The origin of the 200~kHz noise floor seen in the last decade is less certain. From the comb versus comb results discussed in the previous section, the noise floor is not due to calibration uncertainty. Temperature induced drift is largely ruled out when we apply the temperature sensitivity of the lines given in~\cite{Swann2005hcn} to our measured lab temperature shifts. In addition, we also did not find a significant coupling between frequency shifts and variations of the mean LHR amplitude. While unexplored, there could be a polarization induced effect due to the non-polarization maintaining fibers  used to connectorize the HCN system. This is an area that needs further investigation, but regardless of the true source or cause of the noise floor it does not yet represent a concern for the solar measurements we performed. With the current solar tracking setup the obtainable spectral averaging times on the Sun are less than 1000~s, which is still within the $1/\sqrt{\tau}$ averaging region of the HCN dataset.

An estimate of the HCN line's absolute frequency is used to isolate the absolute comb line number ($n$). Since the SRM~2519a HCN reference cell is only certified for P-branch frequencies up to the P27 transition~\cite{srm2519a}, its documentation does not report a frequency specification for the P28 transition. We estimate our cell's P28 frequency by adding the pressure shifts extrapolated from the last few certified lines to the vacuum line center calculated in reference~\cite{Swann2005hcn}. That estimate was found to be within 2~MHz of the frequency given by our LHR measurement and an integer comb number, which results in a comb-calibrated absolute frequency of 191.5114832(2)~THz for the P28 transition of our gas cell. For comparison, a 2~MHz error is well within the ${\sim}10$~MHz $\pm 1 \sigma$ certified uncertainty for the nearest line at P27. The 200~kHz uncertainty, from the Allan deviation shown in Fig.~\ref{fig:HCN}(b), represents an absolute frequency precision of 1~part~in~$10^{9}$.

\section{Results on the Sun}
\label{sec:Sun}

\begin{figure}[tb!]
    \centering
    \includegraphics[width=\figurewidth]{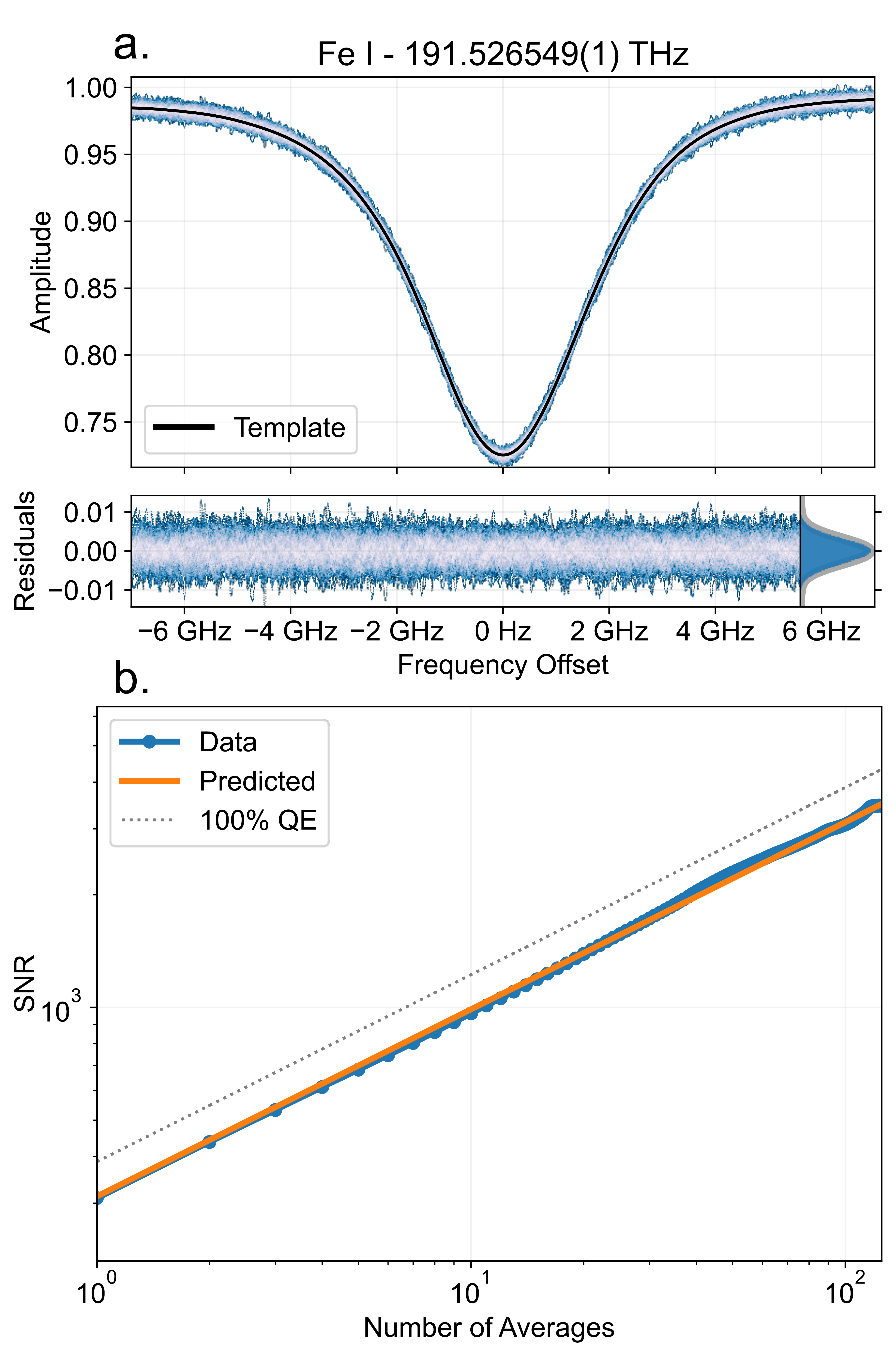}
    \caption{LHR measurement of a neutral iron line in the solar spectrum. The absolute frequency and spectra are given in the frame of the solar system barycenter~\cite{Wright2020bcc}.
    \textbf{(a)}~Fe~I line near 1565.279~nm~\cite{NIST_ASD, Nave1994iron}. The data represents 125, 10~s traces taken over the course of 20~min. The amplitude of the solar spectrum has been normalized to the background continuum and the color is scaled to the density of points. The residuals with respect to the spectral template are plotted below. The inset at the bottom right shows the histogram of the residuals outlined by a Gaussian distribution of the same width in grey.
    \textbf{(b)}~The measured and theoretical SNR of the background continuum with respect to the residuals. Using equation~\ref{eq:SNR}, the predicted single trace SNR is 313 (150~MHz, 21~ms, 6000~K). The data gives an SNR of 309 and increases with $\sqrt{N}$ averaging. The grey dotted line, which starts at 390, represents the predicted SNR assuming the same setup but with no loss and perfect detection efficiency.
}
    \label{fig:IronLine}
\end{figure}

Using this system we observed a neutral iron line (Fe I) in the solar spectrum with a vacuum wavelength near 1565.279~nm~\cite{NIST_ASD, Nave1994iron}. Fig.~\ref{fig:IronLine} illustrates the spectral profile and amplitude noise from an observing run that extended 20~minutes. Each trace, spanning 27.5~GHz, was taken over a 10~s period with 200~MHz resolution bandwidth and 21~ms averaging time. The data was processed using the same procedures as described in the previous section and in the Supplement for the measurement of the HCN line. While the full shape of the iron line only spans ${\sim}$15~GHz, the additional scan range facilitated determination of the comb's integer index $n$ by extending over the positions of both the iron and P28 HCN lines with a single sweep configuration. After the solar measurements finished, the input to the LHR arm was switched to the ASE illuminated HCN gas cell and its known line center was used to determine the reference comb line number to use for absolute frequency calibration of the solar data. While the HCN line is used to determine the integer $n$, we emphasize that the absolute accuracy of the frequency axis is determined by the NIST-calibrated hydrogen maser.

It is interesting to examine the solar data's signal-to-noise ratio as compared to the theoretical value for a thermal source. For this purpose, we take the standard deviation of the residuals over a 4~GHz window roughly 15~GHz away from the line center where the measured signal was closest to the background continuum level~(0.994). With the continuum normalized to 1, the reciprocal of that standard deviation defines the data driven SNR. The predicted SNR is calculated using equation~\ref{eq:SNR}. The total quantum efficiency is 0.77, which accounts for the detector’s quantum efficiency (0.82) and the transmission through the fiber connectors and splitter (0.93). The effective temperature within the field of view of the collimated Gaussian mode (${\sim}25\%$ of the solar disk) is estimated to be 6000~K ($\left< n\right>\! \approx 0.27$), given the nominal solar disk temperature of 5800~K~\cite{Andrej2016sun} and the limb darkening profile at 1565~nm as measured in reference~\cite{Pierce1977limb} (the center of the solar disk is brighter than the average due to limb darkening and so has a larger effective temperature). With the above values, and the system's effective averaging time of 21~ms and equivalent noise bandwidth of 150~MHz, the predicted SNR is 313. This exhibits remarkable agreement with the SNR of 309 calculated purely from the data, and the two values are essentially the same (310) given the numerical precisions involved. Additionally, the data shows the SNR continuing to improve with $\sqrt{N}$ averages at all time scales, reaching a value of approximately 3500 after 125 averages~(20~min). To our knowledge, this is the first demonstration of an LHR measurement that is at the fundamental detection limits dictated by the quantum optical theory of~Eq.~\ref{eq:SNR}.

\begin{figure}[tb!]
    \centering
    \includegraphics[width=\figurewidth]{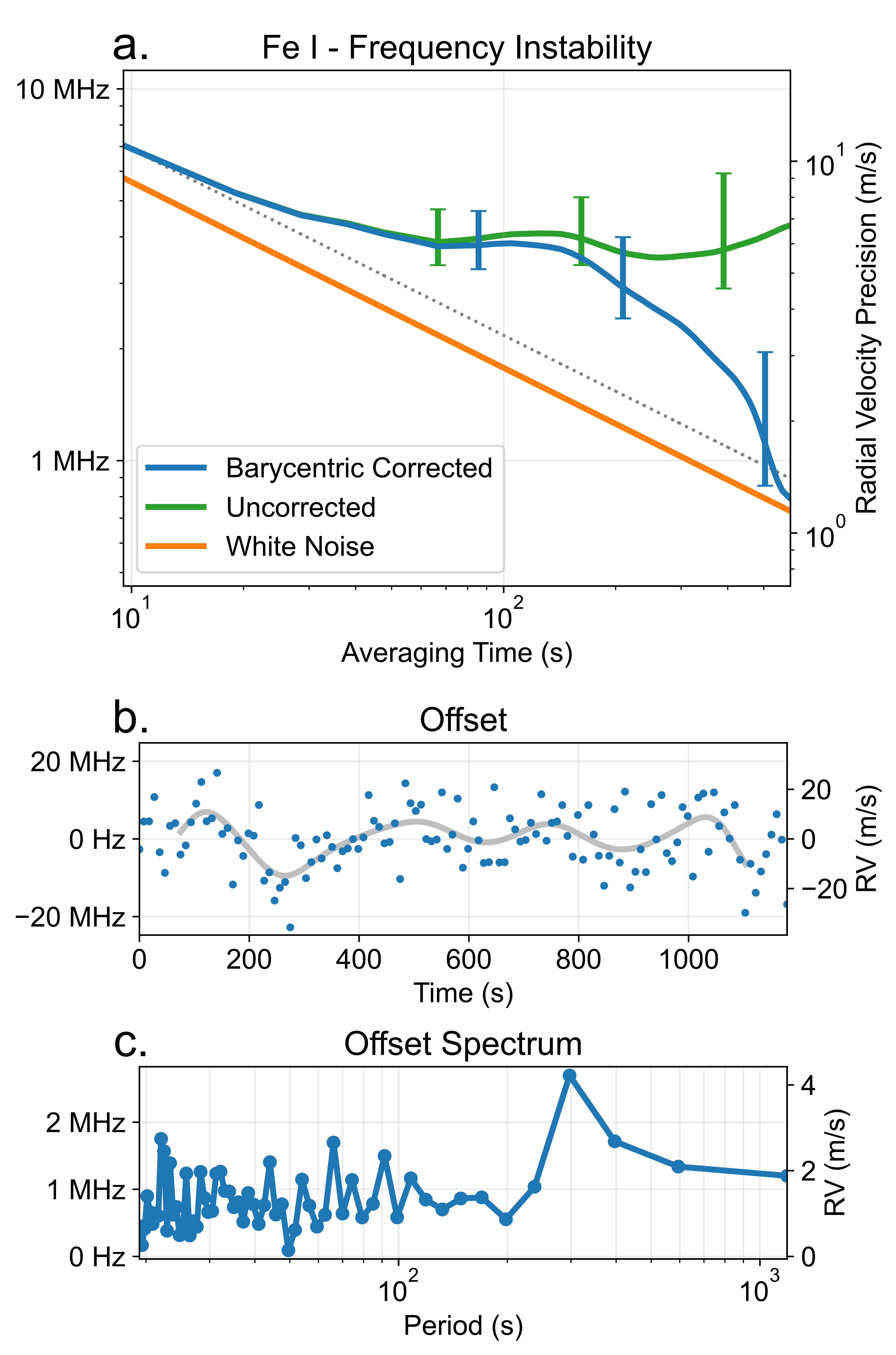}
    \caption{Frequency stability analysis of the iron line in Fig.~\ref{fig:IronLine}.
    \textbf{(a)}~The Allan deviation of the retrieved line centers and the theoretical white-amplitude-noise limit. The error bars represent the $\pm 1 \sigma$ confidence intervals, while the grey dotted line represents $1/\sqrt{\tau}$ averaging and is included to help guide the eye.
    \textbf{(b)}~The frequency offset of the barycentric corrected data in the time domain. The frequency instability is dominated by a low frequency perturbation that occurs at the roughly 5~minute time scale characteristic of solar surface oscillations. The grey line, calculated through a weighted local polynomial filter, reveals those fluctuations to the eye.
    \textbf{(c)}~The spectral amplitudes of the data in part~(b), again demonstrating a prominent peak at about 5 minutes.
}
    \label{fig:FreqStab}
\end{figure}

The central frequency of the line was tracked through cross-correlation with the spectral template, the Allan deviation of which is shown in~Fig.~\ref{fig:FreqStab}(a). The 7~MHz line center instability per 10~s trace represents a measurement precision of 1 part in 500 relative to the 3.7~GHz linewidth. However, because of a variable Doppler shift attributable to the relative motion between the Sun and the LHR system on the surface of the Earth, that value does not significantly improve even after 10~minutes of averaging. Barycentric correction, using the algorithms from references~\cite{kanodia2018barrycorpy, Wright2020bcc}, removes that effect and calculates the expected frequency as seen by an observer far outside the solar system and at rest with respect to the solar systems’ center of mass. Applying that correction, the uncertainty averages down and is reduced below 1~MHz after 10~minutes, a precision which splits the line to better than 1 part in 4000. This is about 40 times more precise than the uncertainties listed for this transition in the NIST Atomic Spectra Database~\cite{NIST_ASD}. The rigorous spectroscopy enabled by such absolute frequency precision should allow investigations into the physics that go into the line frequency and the line shape, including effects such as magnetic activity, convective blueshift, and gravitational redshift~\cite{leone2003measuring, lohner2018convective, hernandez2020solar}. Examining the frequency offsets in Fig.~\ref{fig:FreqStab}(b-c), the short term stability appears to be limited by the p-mode surface oscillations of the Sun, which have characteristic periods around 5 minutes. At timescales longer than shown here, the instability of our solar tracker becomes significant and the precision degrades due to pointing errors asymmetrically coupling the Sun’s large rotational Doppler shift across the system's field of view. We present a photonic solution to this issue below.

\section{Discussion and Future Work}
\label{sec:ideas}

These experiments illustrate the present capabilities of our frequency-comb-calibrated LHR setup for precision spectroscopy, but several important opportunities and open questions remain about the full potential of this approach. The most immediate is the extension of the solar measurements to longer averaging times. Spectral information from across the Sun’s surface is weighted by the antenna pattern given by the distribution of the SMF's optical mode in the far-field~\cite{Siegman1966}. The present results were obtained with optics that coupled light into the SMF from only ${\sim}25\%$ of the solar disk. This leads to solar tracking errors manifesting as spectroscopic noise, as light from the eastern and western limbs of the solar disk have oppositely signed Doppler shifts due to the Sun's rotational motion. There are two options to minimize this effect, reduce the solar tracking errors and/or manipulate the antenna pattern to reduce the sensitivity to misalignment. The first could be accomplished through active feedback that locks to the bright center of the solar disk, while an attractive solution for the second is to employ a flat-top beam shaper.

\begin{figure}[bt!]
    \centering
    \includegraphics[width=\figurewidth]{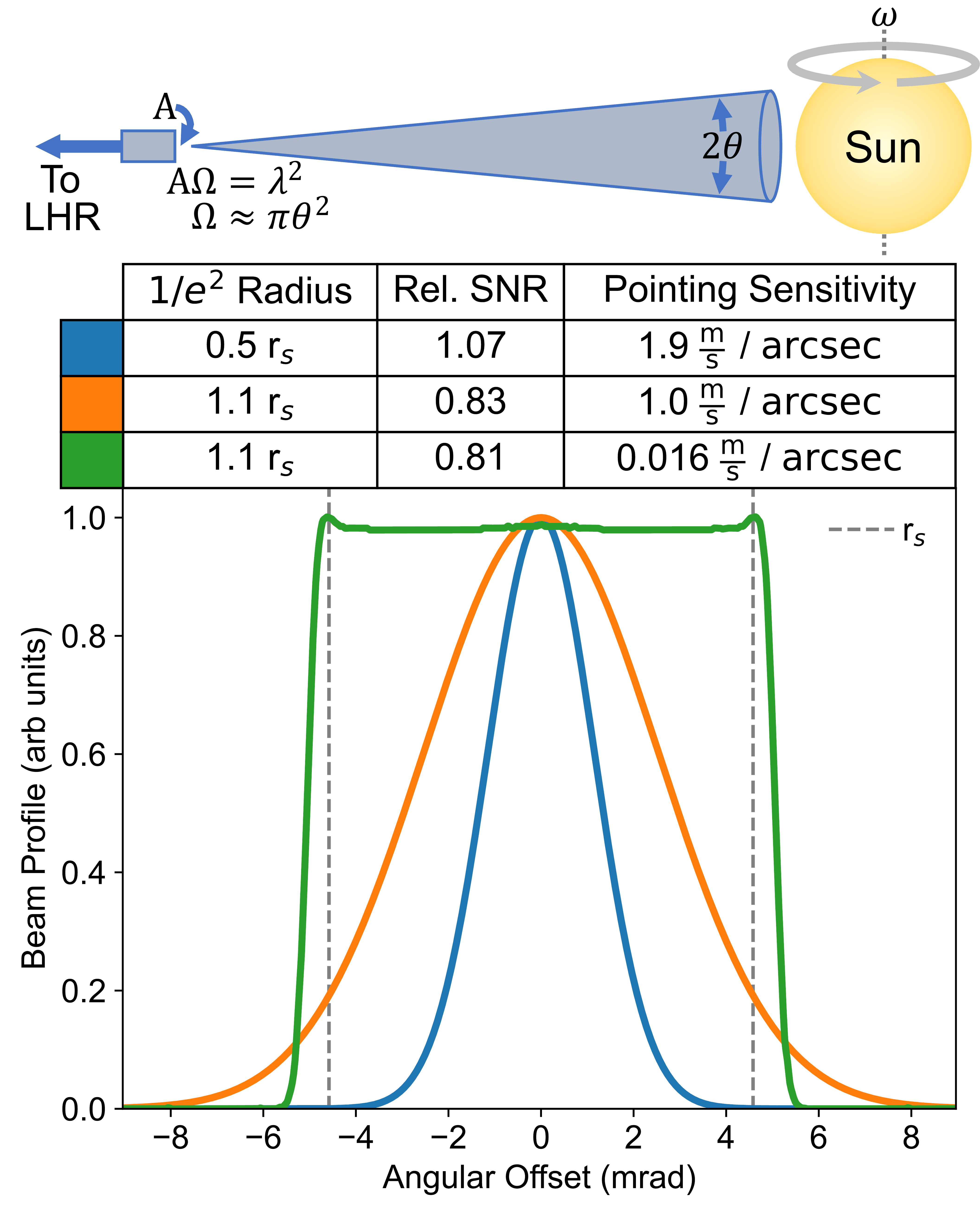}
    \caption{Heterodyne Antenna Pattern and Pointing Sensitivity. \textit{(blue)}~Gaussian mode with radius $0.5$ times the solar radius~($r_s$), \textit{(orange)}~Gaussian mode with radius $1.1 r_s$, and \textit{(green)}~flat-top mode with radius $1.1 r_s$. The relative SNR is normalized to the view given by a uniform antenna pattern that extends up to the edge of the solar disk and includes the effects of limb darkening~\cite{Pierce1977limb}. The pointing sensitivity is the effective Doppler shift (due to the Sun’s rotational motion) per angular misalignment perpendicular to the rotational axis ($1$~arcsec is about $1$ part in $1000$ of $r_s$). The sensitivity of the Gaussian mode does not significantly decrease upon increasing its field of view from $0.5$ to $1.1$ solar radii, but a similarly sized flat-top beam should reduce the pointing sensitivity by 2 orders of magnitude compared to the current $0.5 r_s$ setup.}
    \label{fig:beamprofile}
\end{figure}

In contrast to those based on diffusers, which have been applied to great effect in precision astronomical photometry~\cite{Stefansson2017diffusers}, the beam shapers proposed for use with LHR are single-mode in nature. Such beam shapers are common in laser machining and use refractive or diffractive elements to convert a single-mode Gaussian profile in the near-field into a flat-top profile in the far-field without introducing a speckle pattern. Fig.~\ref{fig:beamprofile}~compares the antenna pattern efficiency and pointing sensitivity of two Gaussian distributions with a flat-top profile provided by the manufacturer Holo/Or. While only a small perturbation for these beam shapes, the pointing sensitivity calculation includes models for limb darkening~\cite{Pierce1977limb} and the latitude variability of the Sun’s rotational velocity~\cite{Snodgrass1990}. In comparison to a Gaussian mode profile, a well matched top-hat aperture reduces the tracking sensitivity by $100\times$. 

Viewing the Sun with a uniform antenna pattern, effectively as if it were unresolved, would allow the lessons learned from such LHR measurements to be more easily transferred to the study of distant stars. For example, it has been shown that radial velocity measurements of the Sun are correlated with its magnetic activity, and that removing those correlations improves long-term precision~\cite{haywood2020unsigned}. However, it is not currently known how to apply those corrections to other stars where precision magnetic field measurements do not exist (such as the spatially resolved measures used in~\cite{haywood2020unsigned}). A potential solution may be to extract the magnetic field information from the shape of the observed spectral lines, which can manifest as line splitting, line broadening, or line intensification~\cite{stiff2003intense}. It is for this purpose that we chose an iron line with a high magnetic sensitivity~\cite{penn2014infrared}. With the experimental flexibility that LHR provides, we are planning to undertake a series of measurements to determine the resolution, signal-to-noise ratio, and temporal sampling required to take full advantage of the stellar activity information contained within the line shape itself. While near-infrared LHR is largely restricted to targeting the Sun (see Supplement), such measurements could still set important constraints on the design of future observations and spectrographic instrumentation for other stars.

On the other hand, with a larger aperture telescope LHR could be used to achieve high-spatial-resolution integral field spectroscopy on the Sun. Sunlight from multiple positions in the solar disk could be coupled into a bundle of single-mode fibers and spatially-resolved spectroscopy could be performed in parallel with a common local oscillator. Spatially and temporally resolved spectra are used to probe the temporal evolution of granulation on the Sun, and to constrain the 3D magnetohydrodynamics code that simulate the solar photosphere. While the newly commissioned Daniel K. Inouye Solar Telescope~(DKIST) can also perform high resolution spectroscopic observations~\cite{rimmele2020dkist, rast2021dkist}, frequency-comb-calibrated LHR could intrinsically provide frequency information with higher fidelity and precision.

An interesting direction for future work is to employ the frequency comb itself as the local oscillator. By channelizing the spectra of the thermal light around each comb tooth, this could significantly broaden the instantaneous spectral bandwidth over which measurements could be made. For solar physics, coverage of several lines would allow simultaneous exploration of a range of magnetic sensitivities and photospheric depths of formation; for spectroscopy of the Earth's atmosphere, this would provide simultaneous information on multiple species (e.g. CH$_4$, CO$_2$, H$_2$O, ...) as well as full molecular rovibrational spectra, which is useful for temperature and pressure quantification.  Preliminary experiments using frequency combs to directly measure thermal-like sources have been performed~\cite{Giorgetta2010, Boudreau2012}. However, in these works the thermal heterodyne signal required extremely long averaging times to emerge above the shot, thermal, and electronic noise floors. A more fruitful direction may be to combine spectral de-multiplexing with combs of high power per mode to optimize the SNR on an array of detectors. We provide a few scenarios in the Supplementary Material.

Focusing on the frequency content of a single line, the achievable frequency precision could be increased by discarding the line shape information. A signal proportional to the derivative of the spectral amplitude, suitable for use as an error signal to servo and lock the LO laser~\cite{Weel2002lockin}, can be generated by dithering the laser frequency and performing lock-in detection of the LHR output. Such derivative signals have been previously demonstrated with LHR~\cite{Martin-Mateos18prop, Martin-Mateos19impl}, but not in the context of feeding back to and stabilizing the laser frequency. Locking the LO laser would make the average laser frequency a direct representation of a line's instantaneous frequency. This technique would maximize the achievable SNR and hence the achievable frequency precision by concentrating the effective averaging time at the point of spectroscopic interest, by allowing the use of an increased resolution bandwidth optimized to the linewidth, and by enabling the line frequency to be continuously tracked with minimal dead-time against the laser frequency comb. For the iron line in Fig.~\ref{fig:IronLine}, we calculate an optimal resolution bandwidth of about 3~GHz based on the theoretical increase in SNR and the decrease in the slope of the line with larger resolution bandwidth. We estimate that with a square-wave-type dither jumping between the two steepest parts of the line, the resulting SNR should support 1~m/s precision in only 10~s and 10~cm/s in 1000~s. Besides being a path towards cm/s radial velocity precision on the Sun using only a single line, such measurements could also complement single-line helioseismic studies such as the BISON and GONG networks~\cite{Chaplin1996bison, Harvey1996gong} by extending high precision radial velocity measurements to different wavelengths and atomic species.

Polarization is an aspect of our work that remains to be analyzed. As heterodyning is a fundamentally single-mode process, the signal only emerges from the thermal light projected onto the polarization state of the LO. The absorption lines embedded in the solar spectrum are slightly polarized due to the Sun's magnetic field, and the Earth’s atmosphere can in principle change the polarization of the detected light randomly at the few percent level~\cite{Plass1970scattering}. While we have yet to see the direct effect of this noise source in our measurements, mitigation could either involve fully randomizing (scrambling) the polarization of the sunlight via agitation/rotation/squeezing of the SMF in which the solar light propagates, or splitting and then separately detecting the two orthogonal polarizations of the input solar light. The second approach would be particularly beneficial for the analysis of magnetically-sensitive lines as it could allow mapping of the full polarization state of the transition, giving detailed information on the local magnetic fields~\cite{iglesias2019instrumentation}.

While challenging, the introduction and continued improvement of photonic tools and optical frequency combs makes further study of LHR in the near and mid-infrared a compelling cross-disciplinary field of exploration for precision spectroscopy and Doppler metrology in astronomy, atmospheric science, and remote sensing.

%%%%%%%%%%%%%%%%%%% Back Matter %%%%%%%%%%%%%%

\begin{backmatter}
\bmsection{Funding}
NIST (70NANB18H006) and NSF (AST-1310875, AST-1310885, AST-2009982, ATI 2009889, ATI 2009982)

\bmsection{Acknowledgments}
The Center for Exoplanets and Habitable Worlds is supported by the Pennsylvania State University, the Eberly College of Science, and the Pennsylvania Space Grant Consortium. The authors further acknowledge helpful comments and discussions on the topic with Eugene Tsao, and Bill Swann.  This work is a contribution of NIST and is not subject to copyright in the US.  Mention of specific products or trade names is for technical and scientific information and does not constitute an endorsement by NIST.

\bmsection{Disclosures} The authors declare no conflicts of interest.

\bmsection{Data availability} Data underlying the results presented in this paper are not publicly available at this time but may be obtained from the authors upon reasonable request.

\bmsection{Supplemental document}
See Supplement 1 for supporting content. 

\end{backmatter}

%%%%%%%%%%%%%%%%%%% References %%%%%%%%%%%%%%%

\bibliography{references}

\end{document}

\maketitle

\thispagestyle{empty}
\section{Spectroscopic Data Processing}
\begin{figure}[b!]
    \centering
    \includegraphics[width=0.75\linewidth]{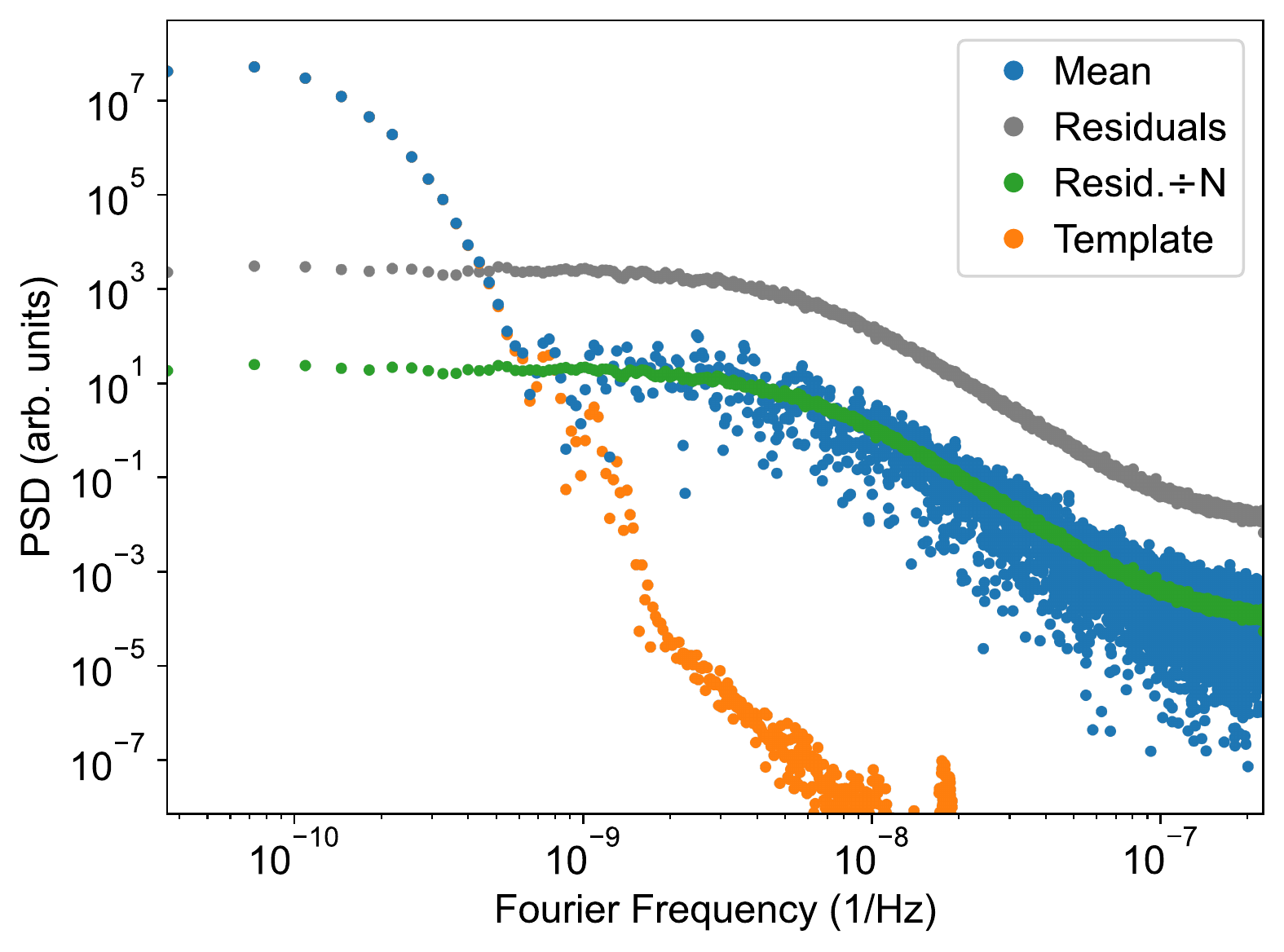}
    \caption{
    Power spectral density (PSD) of the iron line data: \textit{(blue)}~the mean line amplitude, \textit{(grey)}~the residuals between the mean and individual traces, \textit{(green)}~the residuals spectrum divided by the number of traces, and \textit{(orange)}~the line template. All spectra have been processed using a Blackman window. The residuals spectrum represents the average PSD of the residuals from all 125 traces. The average residual spectrum divided by the number of traces is equivalent to the noise floor of the mean line amplitude. The template is formed by filtering out noise at high frequencies, with the bandwidth of the filter chosen to correspond to the point at which the signal emerges from the noise floor. The low frequency noise of the residuals is white but falls off at high frequencies. This roll-off is due to the time-averaging low-pass filter and its effect is included in the estimate of the white-amplitude-noise limit.}
    \label{fig:template}
\end{figure}

The amplitude and frequency noise of the spectroscopic signals (HCN and solar iron line) were measured against a template formed from a low-pass filtered average of the datasets. After frequency axis calibration, the raw voltage signals are first shifted and scaled to represent RF noise power. This is performed using the noise-power-to-voltage transfer function that was measured with an amplified spontaneous emission source. Before being averaged together the traces are scaled to the same mean amplitude to normalize out changes in source brightness, and for the solar data the traces are barycentric corrected~\cite{kanodia2018barrycorpy, Wright2020bcc}. The final template is calculated by removing high frequency noise from the averaged spectrum using a weighted local polynomial filter. The filter is implemented with a 5th order polynomial and a Gaussian weight whose width is chosen through inspection of the noise profile in the line amplitudes' Fourier spectrum (see Fig.~\ref{fig:template}). The average frequency of the templated line and the level of the background continuum are calculated by fitting a Voigt profile to the template using data from within 2 line widths of the line center. The result is used to normalize the continuum level of the template to 1.

The theoretical sensitivity of the extracted line center frequency to white-amplitude noise is estimated by finding the effective line center shift of the template after adding a set amount of Gaussian amplitude noise. Numeric simulations show that this white-noise limit scales linearly with the amplitude noise times the ratio of the line width to the line depth. The distribution of the noise also requires consideration. Since the Fourier transform of cross-correlation is multiplication, significant contributions to the retrieved line center uncertainty only come from amplitude noise at periods on the order of magnitude of the linewidth or greater. In the Fourier domain picture, the line shape of the template acts as a low-pass filter. Since the digitizer collects many samples per effective averaging window the amplitude noise of the measured signal is not equally distributed spectrally (see Fig.~\ref{fig:template}). In order to produce the white noise curve in Fig.~3(b) and 5(a) of the main text, numerically generated white noise is first convolved with a second order exponential filter that has the same effective averaging time as the second order RC filter used in the experiment. The gain of the filter is also scaled such that the noise has the same variance as the experimental residuals.

\section{Achieving Broader Bandwidths with Multiplexed Measurements}

\begin{figure}[b!]
    \centering
    \includegraphics[width=0.75\linewidth]{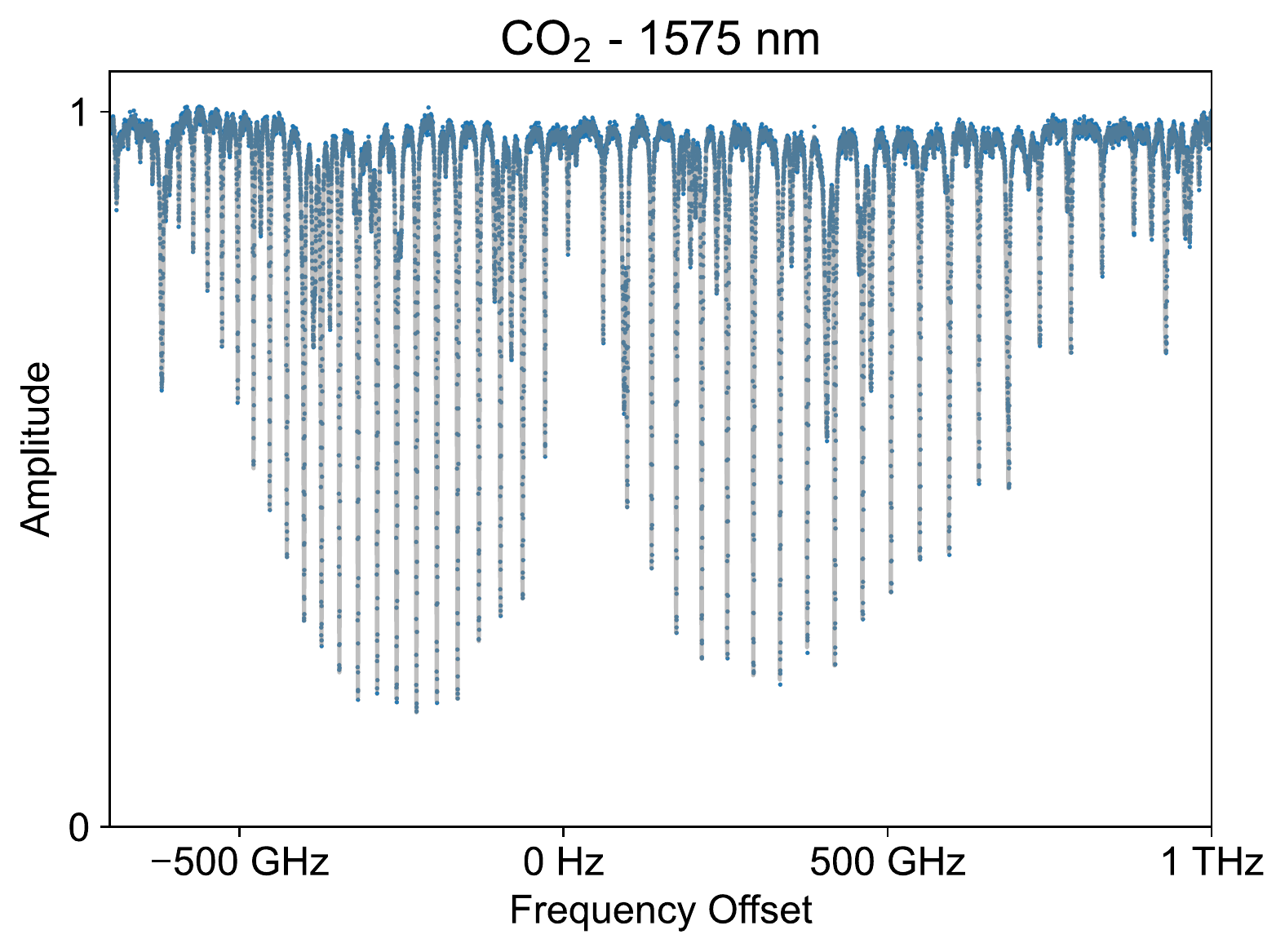}
    \caption{Wideband LHR measurement highlighting the rovibrational band of CO$_2$ near 1575~nm. The scan spans more than 1.5~THz of bandwidth. The data represents the average of 100, 7~s traces taken over the course of 10 minutes with $100$~MHz resolution bandwidth and 0.2~s averaging time. The CO$_2$ lines are due to absorption throughout the entire atmospheric column. Additional spectral features arise from Solar absorption lines.}
    \label{fig:co2}
\end{figure}

Our experiments lay the groundwork for extension of LHR to broader bandwidths. In a straightforward approach, a more broadly tunable laser could accomplish this, and continuous coverage from 1260 nm to nearly 1700 nm is available with commercial diode lasers. Along these lines, we have performed proof-of-concept measurements of the atmospheric CO$_2$ bands near 1575 nm, and results are shown in figure~\ref{fig:co2}. However, due to the trade-off between total optical bandwidth and total scan time, the achievable SNR becomes more and more limited as the scanned bandwidth increases. A more efficient option would be to use multiple local oscillators and simultaneously perform LHR on thermal light that has been separated into spectral channels. This would have the advantage of parallelizing the data acquisition, and with sufficiently low-loss components the measurement would yield a net SNR increase. Ignoring losses, splitting the thermal light into $N$ spectral channels before detection would increase the achievable SNR (for fixed scan time) or decrease the total scan time (for fixed SNR) by a factor of $\sqrt{N}$.

\begin{figure}[tb!]
    \centering
    \includegraphics[width=1\linewidth]{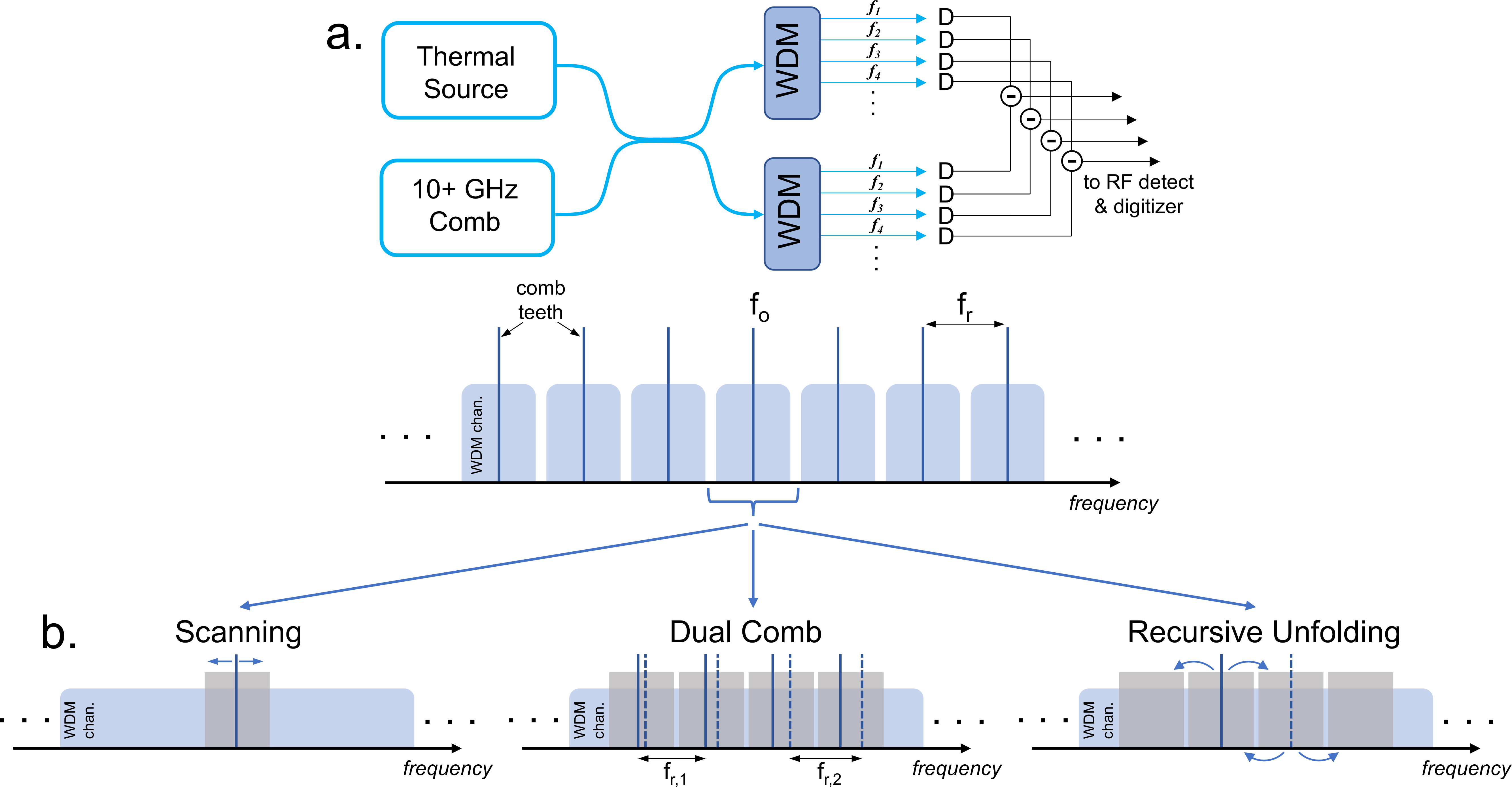}
    \caption{Extending LHR bandwidth through multiplexing.
    \textbf{(a)} After combining light from a thermal source and a high repetition rate frequency comb, wavelength division multiplexers (WDM) separate the light into channels for detection.
    \textbf{(b)} High resolution spectra can be obtained within each WDM channel by scanning the comb teeth, or by replicating the setup and measuring two copies of the thermal light, either through recursive unfolding or dual comb techniques. The grey boxes represent the effective optical bandwidth of the high resolution elements and the dashed lines represent the position of the second frequency comb.}
    \label{fig:multiplex}
\end{figure}

A powerful implementation of parallelization would use the frequency comb itself as the local oscillator. As shown in figure~\ref{fig:multiplex}(a), the entire comb could be combined with the thermal light source in a 50:50 fiber beamsplitter and then spectrally de-multiplexed into channels by wavelength division multiplexers (WDM). The output of each channel would be spaced by the repetition rate of the comb and be sent to its own balanced detector for LHR processing. Such an approach would broaden the spectral coverage by a factor equal to the number of comb teeth, and since the LO's are the known comb lines there would be no need for a separate frequency calibration arm. A comb with a repetition rate on the order of 20-100 GHz might be employed using dense WDMs in the 1.5~$\mu$m telecom region. Such high repetition rate combs can be generated using electro-optics or microresonators, in which a CW laser forms the central tooth of the comb~\cite{Kippenberg2011, Kippenberg2018, Drake2019, Carlson2018, Metcalf2019}. For present electro-optic and microresonator comb systems, the number of teeth can range from several hundred to thousands. The complexity of such multiplexed setups may be reduced by digitizing the heterodyne signal and performing the RF power detection numerically. In addition to broadband high-resolution spectroscopy, massively-parallel heterodyne with frequency combs could also be useful for interferometric imaging in the mid-infrared region~\cite{swenson1986radio}, and such techniques have been proposed in the context of the Planet Formation Imager~\cite{ireland2014dispersed, monnier2018planet, bourdarot2021architecture}.

There are multiple opportunities for ultimately measuring high resolution spectra within the relatively coarse WDM channels. In figure~\ref{fig:multiplex}(b) we envision several modes that would be interesting to explore. In all cases, there will be trade-offs between the achievable SNR, WDM channel width, comb tooth power, and the saturation power of available balanced photodetectors. A full analysis of the unique trade-offs for these methods is out of the scope of this paper, but the basic ideas and system requirements are described below.

The first approach is an extension of the scanning method described in the main paper. The mode-spacing of the comb remains fixed, but the offset frequency is tuned to simultaneously scan all the comb teeth. A scan range of $\pm{1}/{2}f_{r}$ would allow continuous spectral coverage.

The other two methods shown in figure~\ref{fig:multiplex}(b) do not require scanning the laser frequency, but do require splitting the thermal light and sending it into two copies of the detection setup, each with its own frequency comb. The negative and positive sides of the spectrum are folded across the LO during heterodyne and two independent measurements are required in order to break that degeneracy. The first of these is a type of dual comb measurement, in which two combs with slightly different repetition rates are separately heterodyned with the thermal light. The data processing is analogous to autocorrelation spectroscopy and experimental details have been described elsewhere~\cite{Boudreau2012}. One important characteristic is that the resolution is given by the nominal comb spacing, which can now be much smaller than the WDM channel width.

The last method involves a recursive unfolding of the baseband spectra, in which two combs with slightly different offset frequencies (but the same repetition rate or mode spacing) are used to break the folding degeneracy of the other's baseband spectrum. The central spectral regions about each LO are processed the same as for the scanning case and the amplitude in the two central bins are known without ambiguity. However, all successive bins contain spectral information from both sides of the LOs. By recursively subtracting off the known bins from the degenerate bins, the full spectrum can be retrieved out to the bandwidth of the detector. This subtraction requires that the separation between the two LOs be equal to the frequency resolution of the unfolded spectrum. The scanning and dual comb methods only need detector bandwidths out to half of the desired frequency resolution, but the unfolding method requires much higher electronic bandwidth, extending up to half the width of the WDM channel.

\section{Scaling Relationships for LHR of Other Stars}
Finally, the main results of this manuscript were demonstrated using the Sun as the thermal source, and a natural question to ask is whether LHR could be as interesting on other stars~\cite{Sappey2020} or on unresolved objects in general. The largest obstacle is the antenna theorem (${A \, \Omega = \lambda^2}$)~\cite{Siegman1966}, which places imposing constraints on the effective aperture and field of view. For example, Alpha Centauri A (the nearest Sun-like star) would need to be viewed with an unobstructed, diffraction-limited aperture spanning about 60~m in order to get a similar SNR as our results on the Sun, which only used a $\sim$1~mm diameter aperture. For unresolved objects, the maximum single-mode efficiency ($\eta$) scales with the filling factor of the source within the field of view~\cite{Abbas1976}, and this derating can be simply estimated as the squared ratio of the star's angular radius to the angular radius of the collecting aperture's field of view. For cold or unresolved objects $({\left< n \right>} {\ll} 1, \eta {\ll}  1)$, laser heterodyne's signal-to-noise ratio takes a significant hit in comparison to a traditional direct-detection spectrograph, which has an SNR that scales with the square root of the detected thermal photon density~($\sqrt{{\eta} {\left< n \right>}}$ instead of the~${\eta}{\left< n \right>}$ of LHR)~\cite{Zmuidzinas2003}. That being said, we note that there are several telescopes presently planned or under construction (Giant Magellan Telescope~\cite{fanson2018overview}, Thirty Meter Telescope~\cite{sanders2013thirty}, and Extremely Large Telescope~\cite{gilmozzi2007european}) that would have apertures sufficient for the half dozen or so largest angular diameter stars.

% Bibliography
\bibliography{references}